\documentstyle[pre,aps,epsf,multicol]{revtex}
\def\stacksymbols #1#2#3#4{\def\theguybelow{#2}
    \def\verticalposition{\lower#3pt}
    \def\spacingwithinsymbol{\baselineskip0pt\lineskip#4pt}
    \mathrel{\mathpalette\intermediary#1}}
\def\intermediary#1#2{\verticalposition\vbox{\spacingwithinsymbol
      \everycr={}\tabskip0pt
      \halign{$\mathsurround0pt#1\hfil##\hfil$\crcr#2\crcr
               \theguybelow\crcr}}}
\def\lapproxeq{\stacksymbols{<}{\sim}{2.5}{.2}}
\def\gapproxeq{\stacksymbols{>}{\sim}{3}{.5}}
\begin{document}
\draft
\title{Anisotropic Interface Depinning -- Numerical Results}
\author{Heiko Leschhorn\footnote{present address: Theoretische Physik III,
Heinrich-Heine-Universit\"at D\"usseldorf, D-40225 D\"usseldorf, Germany}}
\address{
Center for Polymer Studies and Department of Physics, \\
Boston University, Boston, Massachusetts 02215, USA}
\date{\today}
\maketitle
\begin{abstract}
We study numerically a stochastic differential equation 
describing an interface 
driven along the hard direction of an anisotropic random
medium.  The interface is subject to a homogeneous 
driving force, random pinning forces and the surface 
tension. In addition, a nonlinear term due to the anisotropy 
of the medium is included. The critical exponents characterizing the 
depinning transition are determined numerically 
for a one-dimensional interface. The results  
are the same, within errors, as those of the ``Directed Percolation Depinning'' 
(DPD) model. We therefore expect
that the critical exponents of the stochastic differential equation
are exactly given by the exponents obtained by a mapping of  
the DPD model to directed percolation. We find that a moving interface
near the depinning transition is
not self-affine and shows a behavior similar to the DPD model.
\end{abstract}

\begin{multicols}{2}\narrowtext

\section{Introduction}

The driven viscous motion of an elastic interface in a medium
with random pinning forces  
is relevant for the understanding
of various problems in condensed matter physics \cite{BarSta}.
Examples include 
the ordering dynamics of an impure Ising magnet after 
a quench below the critical temperature \cite{NatRuj},
wetting immiscible displacement of one fluid by another in a porous
medium \cite{experi}, and
pinning of flux lines in type-II-superconductors \cite{FisFis,ErtKar}.
In recent years, studies of 
fluid imbibition in paper \cite{Buldyrev,HorSta} and 
of flameless paper burning \cite{ZhaZha}
have been carried out, and observable interfaces allow a direct 
comparison with theoretical predictions.

Common to all of these problems is a competition between
smoothening due to the surface tension and roughening due
to the interaction with the random pinning forces of the 
medium. Further, there
is a competition between the driving force and the
pinning forces, resulting in a depinning transition. 

On a coarse-grained level, it is expected that the dynamics 
of the interface can be described by the following 
continuum equation of motion,
\begin{equation}
\label{v_n}
v_n (\vec r,t)= \gamma K(\vec r)+ F + \eta (\vec r)+
\vec n \cdot \vec \nabla V (\vec r).
\end{equation}
Here, $v_n (\vec r,t)$ is the normal velocity of the interface at 
position $\vec r$. 
The surface tension generates a term proportional to the 
total curvature
$K(\vec r)= -\vec \nabla \cdot \vec n$, where 
$\vec n$ is the normal vector on the interface at 
$\vec r$. The coefficient $\gamma$ measures the stiffness of the 
interface. $F$ is a homogeneous driving force. The last two terms,
$\eta (\vec r) $
and $\vec n \cdot \vec \nabla V (\vec r)$ represent random-field 
and random-bond disorder, respectively. 
The random forces $\eta (\vec r)$ and the random potential
$V(\vec r)$ are short-range correlated in space.

Equation (\ref{v_n}) is considerably simplified by restricting
ourselves to an almost planar interface without overhangs.
A coordinate system $\vec r = ({\bf x},h)$ can be introduced, so that 
the interface position is given by a single-valued function $h({\bf x},t)$. 
The dimension of ${\bf x}$ is denoted by $d$.
Equation (\ref{v_n}) becomes \cite{TanKar}
$$
{1 \over \sqrt g} {\partial h ({\bf x},t) \over \partial t} =
{\gamma \over  g^{3/2}} \nabla ^2h+F + \eta ({\bf x}, h)+
$$
\begin{equation}
\label{ght}
+{1 \over \sqrt g} \left [ \nabla h \nabla V ({\bf x},h) - 
{\partial V ({\bf x},h) \over \partial h} \right ],
\end{equation}

where $g = 1 + (\nabla h)^2$. 

For sufficiently large values of the driving force $F$, the interface 
grows continuously. However, for smaller values of $F$, 
growth on some regions of the interface can come to a halt, due to
the interaction with the quenched disorder.
We say that these regions of the interface have become {\it pinned}.
As the rest of the interface continues to grow, the pinned 
regions can be dragged over the pinning barriers 
by neighboring moving regions. Then, the 
formerly pinned regions advance quickly, 
which can be considered as an avalanche
\cite{BarBul,NarFisint,snep,PacMas}.

The maximum linear size, $\xi$, of the pinned regions 
diverges when $F$ approaches its
critical value $F_c$, 
\begin{equation}
\label{xi}
\xi \sim |F-F_c|^{-\nu}.
\end{equation}
The threshold $F_c$ is the critical point
of a dynamical phase transition, and $\xi $ the corresponding
correlation length.
The role of the order parameter is played by the mean velocity,
$v = \lim _{t \to \infty,L \to \infty} \overline{\partial h / \partial t}$.
($L$ is the system size and the overbar denotes the spatial average over 
${\bf x}$.) The velocity is zero for $F < F_c$, and increases as 
\begin{equation}
\label{v}
v \sim (F-F_c)^{\theta}
\end{equation}
for $F \gapproxeq F_c$. 
On length scales $l \gg \xi$ pinning 
can be neglected and we can therefore replace the argument $h$ in the
disorder terms $\eta({\bf x},h)$ and $V({\bf x},h)$ by $vt$, i.e., the 
quenched disorder crosses over to thermal noise
\cite{NSTL}. Then, the interface is governed by the 
Kardar-Parisi-Zhang (KPZ) equation \cite{KPZ}. 
In this paper, we are interested in the critical behavior on 
length scales $l \ll \xi$, especially when $\xi \to \infty$
at the depinning transition. 

The global interface width 
$w^2 = \left \langle \overline {h ({\bf x},t) - \overline h (t)}
\right \rangle $ is another characteristic quantity of the interface.
Here and elsewhere, $\langle \cdot \rangle $ denotes an average over
the disorder distribution.
Choosing a flat interface as the initial condition,
$h({\bf x},t=0) \equiv 0$, $w^2$  
scales as \cite{FamVic,pdp}
\begin{equation}
\label{w^2sca}
w^2 (\xi,t) \sim \xi^{2\alpha} \Psi _\pm (t/\xi ^z),
\end{equation}
for a sufficiently large system size, $L > \xi$.
If $L<\xi$, the correlation length $\xi$ in Eq. (\ref{w^2sca})
has to be replaced by $L$.
$\Psi _+ (y)$ and $\Psi _-(y)$ 
are scaling functions for $F>F_c$ and $F<F_c$, respectively.
Both functions scale as $\Psi _\pm (y) \sim y^{2 \beta}$
for $y \ll 1$, where $\beta = \alpha /z$. It follows
\begin{equation}
\label{w2t}
w^2 (t) \sim t^{2 \beta}  ~~~~~(t \ll \xi ^z). 
\end{equation}
For $F>F_c$, pinning is irrelevant on length scales 
$l \gg \xi$, so we can neglect
pinning also on time scales $t \gg \xi ^z$. Thus, 
$\Psi_+ (y) \sim y^{2 \beta _m}$  for $y \gg 1$, 
where $\beta _m $ is the growth exponent of an interface 
subject to thermal noise. 
Below threshold, the interface becomes pinned  and
$\Psi _- (y) = const. $
Using the scaling of $\Psi_{\pm} (y)$ for $y \ll 1$ and Eq. (\ref{w^2sca}) 
we obtain
\begin{equation}
\label{w^2cas}
w^2 (\xi,t) \sim \cases{t^{2 \beta_m}  \, &  ($t \gg \xi ^z$, $F > F_c$)\cr
                      \xi ^{2\alpha}\, &  ($t \gg \xi ^z$, $F < F_c$).\cr }
\end{equation}
It has been shown that 
the critical exponents fulfill an exact scaling 
relation \cite{NSTL},
\begin{equation}
\label{exasca}
\theta = \nu(z-\alpha).
\end{equation}

\section{The model for anisotropic depinning}

To further simplify the equation of motion (\ref{ght}), we assume that 
the typical gradients $\nabla h$ are small
on large length scales, so that the roughness exponent $\alpha $ 
is smaller than one. 
This assumption has to be compared with the final results.
When expanding  
$ 1/\sqrt{g} \simeq 1 - (\nabla h)^2/2$, 
nonlinear terms proportional to $(\nabla h)^2$ are generated 
in Eq. (\ref{ght}). A natural question is whether these terms are 
relevant at the depinning transition. 
A term $(\nabla h)^2 $ with a positive coefficient on the 
right hand side of 
Eq. (\ref{ght}) would give a nonzero 
contribution to the driving force for any rough interface. 
This contribution increases when imposing a global tilt of the interface.  
Thus, the threshold $F_c$ becomes a function of the average 
orientation of the interface. 
This is reasonable for anisotropic systems but not for isotropic ones. 
For interfaces in an isotropic environment, it can indeed be shown 
that the nonlinear terms generated by 
expanding $1/\sqrt{g}$ in Eq. (\ref{ght})
are irrelevant close to the 
depinning transition \cite{NarFisint}.
If however, the medium is anisotropic, a term of the form
$\lambda (\nabla h)^2$ can be relevant even for the case $v \to 0$.  
In fact, Tang, Kardar, and Dhar \cite{TanKar} argued that 
if $\nabla V$ and $\partial V / \partial h$ are 
differently distributed, the term $\lambda (\nabla h)^2$ 
is generated under coarse graining.
More generally, if the system is anisotropic in the sense
that the threshold $F_c$ depends on the average orientation 
of the interface, $\lambda (\nabla h)^2$ 
is the only relevant term that can change the universality class
of the depinning transition \cite{TanKar}. Motivated by these observations,
we consider the following equation of motion,
\begin{equation}
\label{dht}
{\partial h({\bf x},t) \over \partial t} = \gamma
\nabla ^2 h + \lambda (\nabla h)^2 + F + \eta ({\bf x}, h).
\end{equation}
For $\lambda > 0$ the threshold $F_c$ has a maximum for an interface 
without tilt, i.e., an interface with periodic boundary conditions 
in Eq. (\ref{dht}) is driven along the hard 
direction of the anisotropic medium \cite{TanKar}.

Equation (\ref{dht}) is the model we study in this paper. 
For simplicity, we restrict ourselves to interface dimension
$d=1$ and 
consider only random-field disorder $\eta (x,h)$. 
It was shown by Narayan and Fisher \cite{NarFisint}
for isotropic systems that random fields and random bonds 
give rise to the same critical behavior. This has been
supported by numerical simulations of interfaces subject to 
random-bond disorder \cite{DonMar,Jensen}.
The random forces are assumed to have zero mean 
and short-range correlations,
$\langle \eta(x',h')\eta(x'+x,h'+h) \rangle = \delta(x) \Delta(h)$, where
$\Delta (h)$ decreases exponentially for $|h|$ greater 
than a microscopic cutoff.

\subsection*{Previous Results}

The case $\lambda = 0$ in Eq. (\ref{dht}) was first investigated by
Feigel'man \cite{Feigelman}. Significant progress
has been made by a functional renormalization-group treatment 
\cite{NarFisint,NSTL,PhD}
and by extensive numerical simulations 
\cite{PacMas,PhD,gimo,RouHan,MakAma,OlaPro}.

The results for the anisotropic
case $\lambda > 0$ are less well established. 
It was first suggested in Ref. \cite{comsne}  
that Eq. (\ref{dht}) is in the same universality class 
as the ``Directed Percolation Depinning'' 
(DPD) model \cite{Buldyrev,pdp}. For $d=1$, 
directed percolating paths of pinning sites stop the interface.
Thus, the roughness of the pinned interface is given by the 
scaling of the directed paths  and the corresponding 
roughness exponent is $\alpha \simeq 0.633$ \cite{Buldyrev,pdp}. 
The other critical exponents of the DPD model
can also be obtained by a mapping to directed percolation; in $d=1$  
the dynamical exponent $z=1$, the correlation  
length exponent $\nu \simeq 1.733$, and the velocity exponent
$\theta \simeq 0.636$.

Amaral, Barab\'asi, and Stanley \cite{AmaBar} measured 
the tilt dependence of the velocity for several versions 
of the DPD model and found a behavior 
consistent with the relevance of the term $\lambda (\nabla h)^2$
at the depinning transition.
Galluccio and Zhang \cite{GalZha} simulated a self-organized 
version \cite{HavBar,Sneppen} of Eq. (\ref{dht}) and obtained
$\alpha \simeq 0.63$, thereby supporting the conjecture of 
Ref. \cite{comsne}.
Recently, Olami, Procaccia, and Zeitak \cite{OlaPro} argued
that the slopes of the pinned surfaces in Eq. (\ref{dht})
are bounded 
and that therefore Eq. (\ref{dht}) could belong to the directed percolation
universality class. 

However, Csah\'ok et al. \cite{Csahok} performed the 
only direct simulation of the 
continuum equation (\ref{dht}) and came to a different conclusion:
Their numerical values for the exponents $\alpha$ and $\beta$ in $d=1$ are 
in agreement with a scaling theory for Eq. (\ref{dht}) which yields
$\alpha = (4-d)/4$ and $\beta=(4-d)/(4+d)$ \cite{Csahok}.
Yet another proposal was made by Parisi \cite{Parisi}. 
He argued that $\beta = (4-d)/4 $ and supported this value in 
$d=1$ by a simulation of a lattice model, which was assumed to be
in the universality class of Eq. (\ref{dht}).
Problematic is also the interpretation of
Stepanow's renormalization-group calculation 
in $d=4-\epsilon$ dimensions \cite{Stepanow}.
The extrapolation to $d=1$ gives the results 
$\alpha \simeq 0.86$, $z \simeq 1$,
$\nu \simeq 1.2$, and $\theta \simeq 0.16$
\cite{Stepanow}.  

\subsection*{Aim of the Paper}

To resolve these discrepancies we carefully determine in this paper
the critical exponents 
by large-scale simulations of Eq. (\ref{dht}) for interface 
dimension $d=1$. 
To this end we carry out a numerical integration of the equation of
motion with a continuous height variable as well as a simulation of
an automaton model where the height variable takes integer values only.
In addition to the previously measured exponents
$\alpha$ and  $\beta$ we also determine the exponents $\nu$ and $\theta$
to strengthen our conclusions. We find that the numerical values
for all critical exponents are in agreement with the suggestion that
Eq. (\ref{dht}) and the DPD model belong to the same universality class.

In addition, we investigate the interface roughness
for $F>F_c$. It is shown that a moving interface is not self-affine
and that the behavior of the roughness is very similar to that of 
the DPD model.

\section{Numerical methods}

\subsection{Continuum equation}

First we simulate Eq. (\ref{dht}) with a discretization of the
transverse coordinate only, $x \to i,~h(x,t) \to h_i(t)$ 
(with $1 \le i \le L$).
The random forces $\eta _i (h)$ are chosen as follows:
Each integer position $h_i$ on a square lattice
is assigned a random number $\eta$ between zero and one.
For non-integer $h_i$ the forces $\eta _i (h)$ are obtained
by linear interpolation \cite{Parisi}.
Finally, the $h$-coordinates of $\eta _i (h)$
in each column $i$
are shifted by a random amount $0 \le s_i < 1$, i.e.,
$\eta _i(h) \to \eta _i (h+s_i)$ \cite{mft}.

At $t=0$ the interface is flat, $h_i(t=0) \equiv 0$.
The interface configuration at $t+\Delta t$ is calculated
simultaneously for all $i$ using the method of finite 
differences \cite{GZcom},

$$
h_i (t +\Delta t) = h_i (t) + \Delta t \Bigl \{
\gamma \big [h_{i+1} (t) + h_{i-1}(t)  - 2h_i (t) \big] + 
$$
\begin{equation}
\label{hdt}
+\lambda \bigl [h_{i+1} (t) - h_{i-1}(t) \bigr]^2 + 
g \eta _i (h_i) \Bigr \} .
\end{equation}
Periodic boundary conditions are used and
$g$ is a parameter measuring the strength of the disorder. 
We choose the parameters  
$\gamma = 5$, $\lambda =1 $, $g=3$ \cite{parameter}
and use $\Delta t = 0.04$, for which 
Eq. (\ref{hdt}) is found to be stable. 
We checked that simulations with 
$\Delta t = 0.01$ yield consistent results.

\subsection{Automaton model}

Since the simulations of the continuum equation (\ref{hdt}) are
computationally expensive, 
we also study a lattice model \cite{gimo} of probabilistic cellular
automata, which allows to determine 
the critical exponents more effectively.

The automaton model is defined on a square lattice
where each cell $[i,h]$ (with $1 \le i \le L$) is assigned a random
force $ \eta _{i,h}$ which takes the value 1 with probability $p$
and $ \eta _{i,h} = -1 $ with probability $1-p$.
During the motion at a given time $t$ the local force
$$
f_i (t) = 
\gamma \big [h_{i+1} (t) + h_{i-1}(t)  - 2h_i (t) \big] + 
$$
\begin{equation}
\label{f_i}
+\lambda \bigl [h_{i+1} (t) - h_{i-1}(t) \bigr]^2 + 
g \eta _{i,h_i} 
\end{equation}
is determined for all $i$.
The interface configuration is then updated  simultaneously for all $i$
\cite{gimo}:
\begin{eqnarray}
h_i (t+1) &=& h_i (t) +1 ~~~~~~~ {\rm if} ~~f_i > 0 \nonumber  \\
h_i (t+1) &=& h_i (t) ~~~~~~~~~~~~{\rm otherwise.}
\label{h_i}
\end{eqnarray}

The difference $p-(1-p)=2p-1$ determines the driving force.
For simplicity, we use the density $p$ as the tunable parameter,
and the depinning threshold is denoted by $p_c$.
The results shown in this paper were performed with the following 
parameters, $\gamma = 10$, $\lambda =1$, and $g=20$ \cite{parameter}.

The growth rule specified by Eqs. (\ref{f_i}) and (\ref{h_i})
can be derived from the continuum equation (\ref{dht}) by 
temporal and spatial discretizations.
In addition,
a simple two-state random force $\eta_{i,h} =\pm 1$
is used.
The discretization implies that the critical slowing down
close to the threshold is reduced. This can be seen as follows.
For $v \to 0 $, the local force for most of the interface 
elements $h_i (t)$ in Eq. (\ref{hdt}) 
is almost zero. Nonetheless, in the finite difference 
approximation Eq. (\ref{hdt}), all $h_i(t) $ and $\eta _{i,h_i}$ have
to be updated at each time step. On the other hand, only a subset 
of values $h_i(t)$ and $\eta_{i,h_i}$ are updated at each time step
in the cellular automata model.

\section{Numerical results}

\subsection{Roughness at threshold}
In this section we determine the critical exponents $\alpha$ 
and $\beta$, defined in Eqs. (\ref{w^2sca}) and (\ref{w2t}), respectively. 
First we find the threshold value $F_c$ and $p_c$ from 
a measurement of the   
interface width $w^2(t)$ for different
values of the driving force. Since the correlation length increases when
$F \to F_c$, the range of the scaling regime 
$w^2(t) \sim t^{2 \beta}$ also 
increases (see Eq. (\ref{w2t})). 
The threshold is estimated as the value where  
the power-law scaling holds for the longest time interval.
This method allows to determine the threshold very accurately.

\begin{figure}
\epsfxsize=\linewidth
\epsfysize=0.275\textheight
\epsfbox{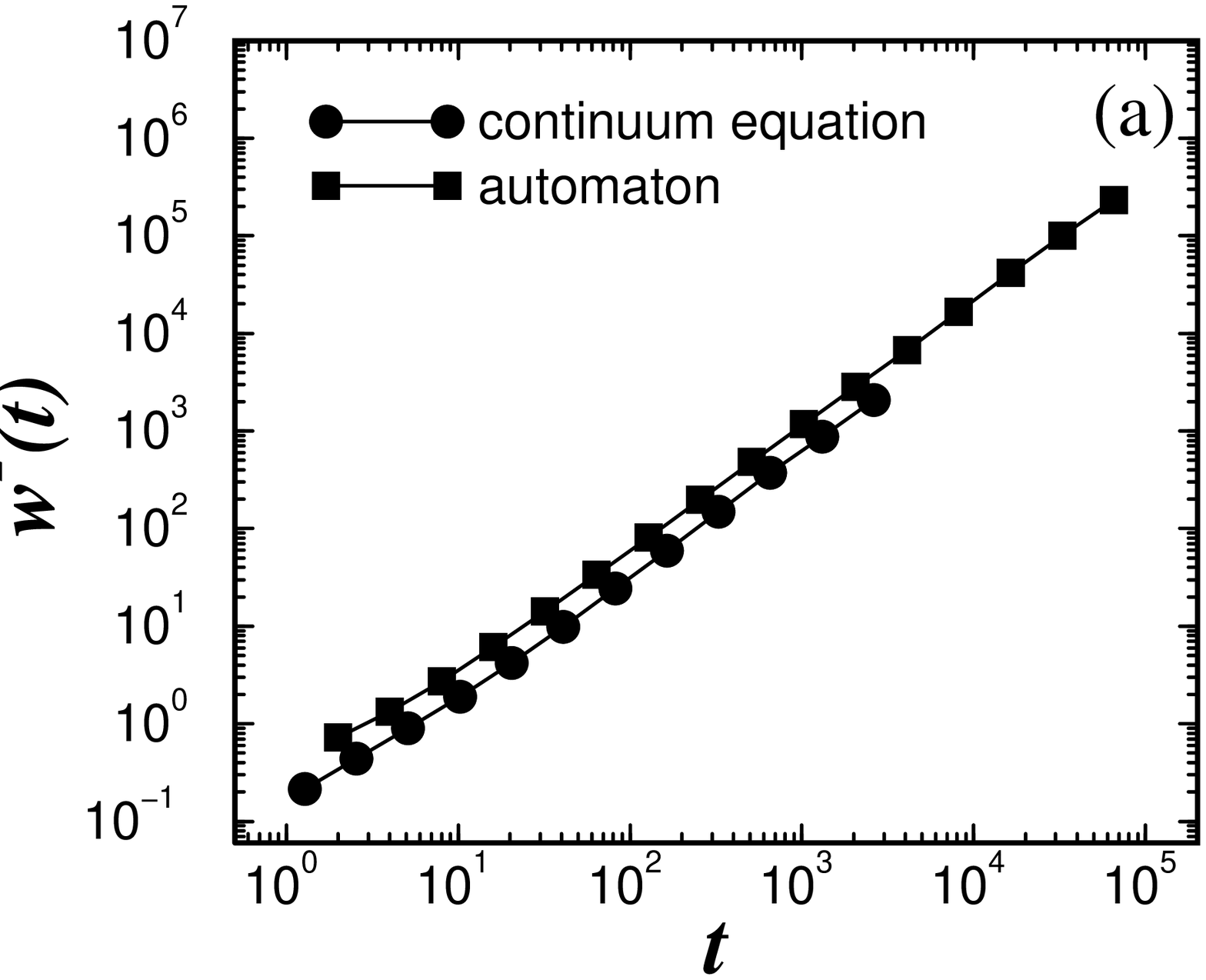}
\epsfxsize=\linewidth
\epsfysize=0.275\textheight
\epsfbox{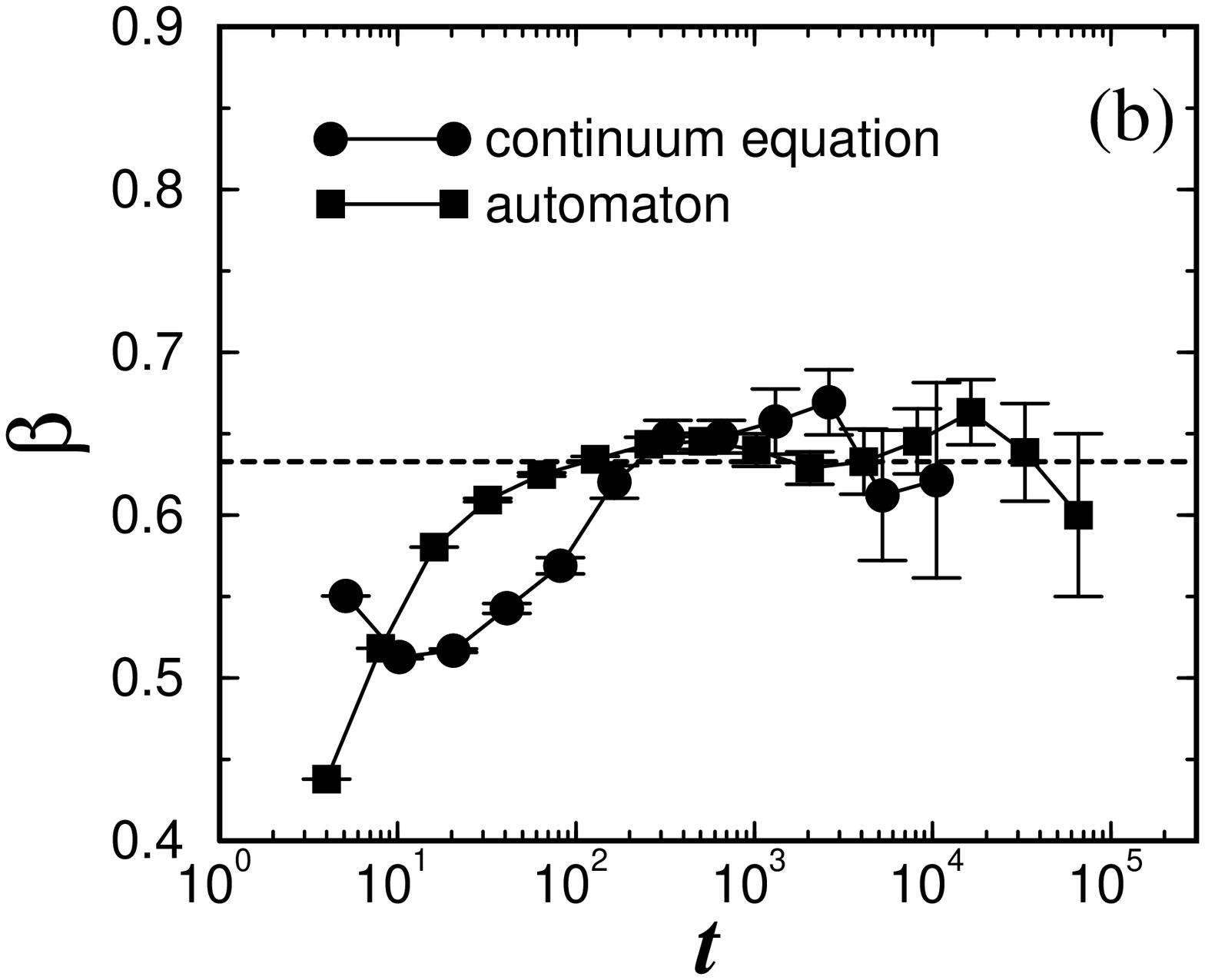}
\caption{
(a) 
Scaling of the interface width $w^2(t) \sim t^{2\beta}$ at 
threshold. 
Interfaces of size $L=16384$
at $F_c \simeq 0.1511$ 
were simulated for the continuum equation (10), 
and at $p_c \simeq 0.6631$ for the automaton ($L=262144$). 
The data were averaged over 40 independent disorder distributions.
The statistical uncertainties are smaller than the size
of the symbols.
The asymptotic slopes yield $\beta = 0.64 \pm 0.02$ for both lines. 
(b)
The corresponding effective exponents 
$\beta (t) = \log [w^2(t)/w^2(t/2)]/\log 4$. 
The value $\beta \simeq 0.633$
of the DPD model is shown as the dashed horizontal line. 
}
\label{wt}
\end{figure}

In Fig. \ref{wt}a the width $w^2(t)$ at the estimated threshold 
is shown in the transient regime, $t \ll \xi$.
In Fig. \ref{wt}b the corresponding effective exponents 
$\beta (t) = \log [w^2(t)/w^2(t/2)]/\log 4$ are shown.
Plotting the effective exponents $\beta (t)$ sensitively shows the 
quality of the power-law scaling. 
{} From Fig. \ref{wt} we conclude that $\beta = 0.64 \pm 0.02$ for
both the continuum equation (\ref{hdt}) and the
automaton (Eqs. (\ref{f_i}), (\ref{h_i})).
This is in very good agreement with the value $\beta \simeq 0.633$,
obtained by mapping the DPD model to directed percolation
\cite{Buldyrev,pdp}.

A convenient estimate of the roughness exponent $\alpha$ 
can be obtained by measuring
the equal-time correlation function
$ C^2(r,t)=\langle \overline {[h_{i+r}(t) -  h_i (t)]^2}   \rangle$
of pinned interfaces.
In Fig. \ref{Cr}a, the scaling $C^2(r) \sim r^{2\alpha}$ for 
$r \ll \xi$ and $F \lapproxeq F_c$ is shown. 
The corresponding effective exponents 
$\alpha (r) = \log [C^2(r)/C^2(r/2)]/\log 4$ are plotted in 
Fig. \ref{Cr}b and \ref{Cr}c.
Equation (\ref{hdt}) and the automaton
yield the same result, $\alpha = 0.63 \pm 0.01$, again in 
agreement with the prediction of directed percolation,
$\alpha \simeq 0.633$.
Using the numerical result for $\beta$, we see that the simulations are 
also consistent with $z=\alpha / \beta = 1$ \cite{Buldyrev,pdp}.

\subsection{Roughness above threshold}

In this section we consider the steady-state behavior of the roughness
for moving interfaces at driving forces $F > F_c$ such that $\xi < L$.
We are interested in the large-time limit, $t \gg \xi^z$,
when the instantaneous velocity of the interface 
fluctuates around its mean value $v$.

Recently, it was proposed that 
the roughness of moving interfaces in the DPD model 
exhibits scaling  
with exponents different from the critical ones \cite{MakAma}. 
This is in disagreement with 
Ref. \cite{pdp}, where it was argued that on scales $l \le \xi$,
the moving interfaces are {\it not} self-affine, because they consist
of pinned regions with $\alpha \simeq 0.63$ and laterally moving
regions with roughly linear slopes ($\alpha \simeq 1$). 
As a consequence, different 
moments of the equal-time correlation function,
$ C^q(r,t)=\langle \overline {|h_{i+r}(t) -  h_i (t)|^q}   \rangle$,
yield different effective roughness exponents, 
$\alpha _q (r) = \log [C^q(r)/C^q(r/2)]/[q \log 2]$. 

\begin{figure}
\epsfxsize=\linewidth
\epsfysize=0.239 \textheight
\epsfbox{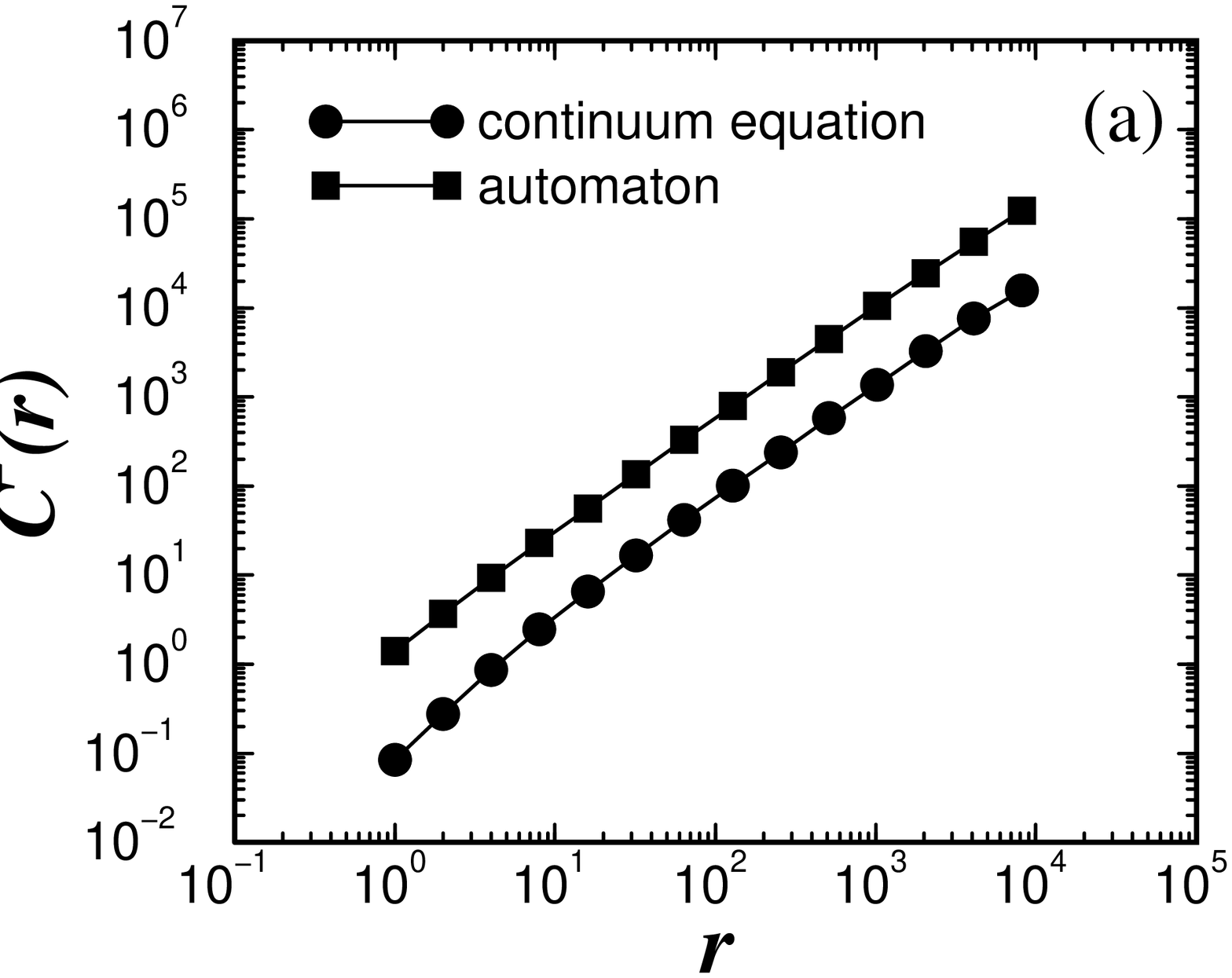}
\epsfxsize=\linewidth
\epsfysize=0.239 \textheight
\epsfbox{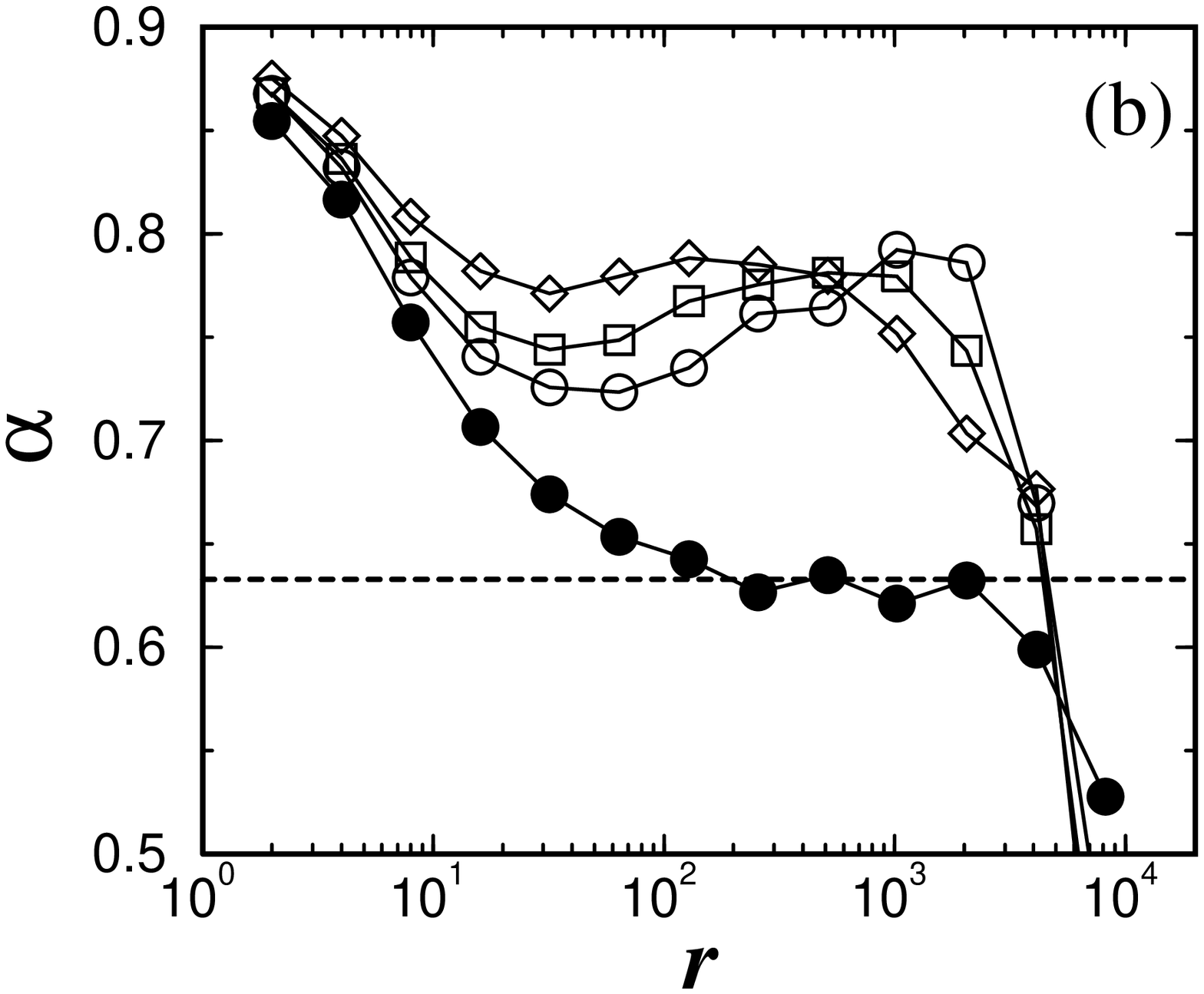}
\epsfxsize=\linewidth
\epsfysize=0.239 \textheight
\epsfbox{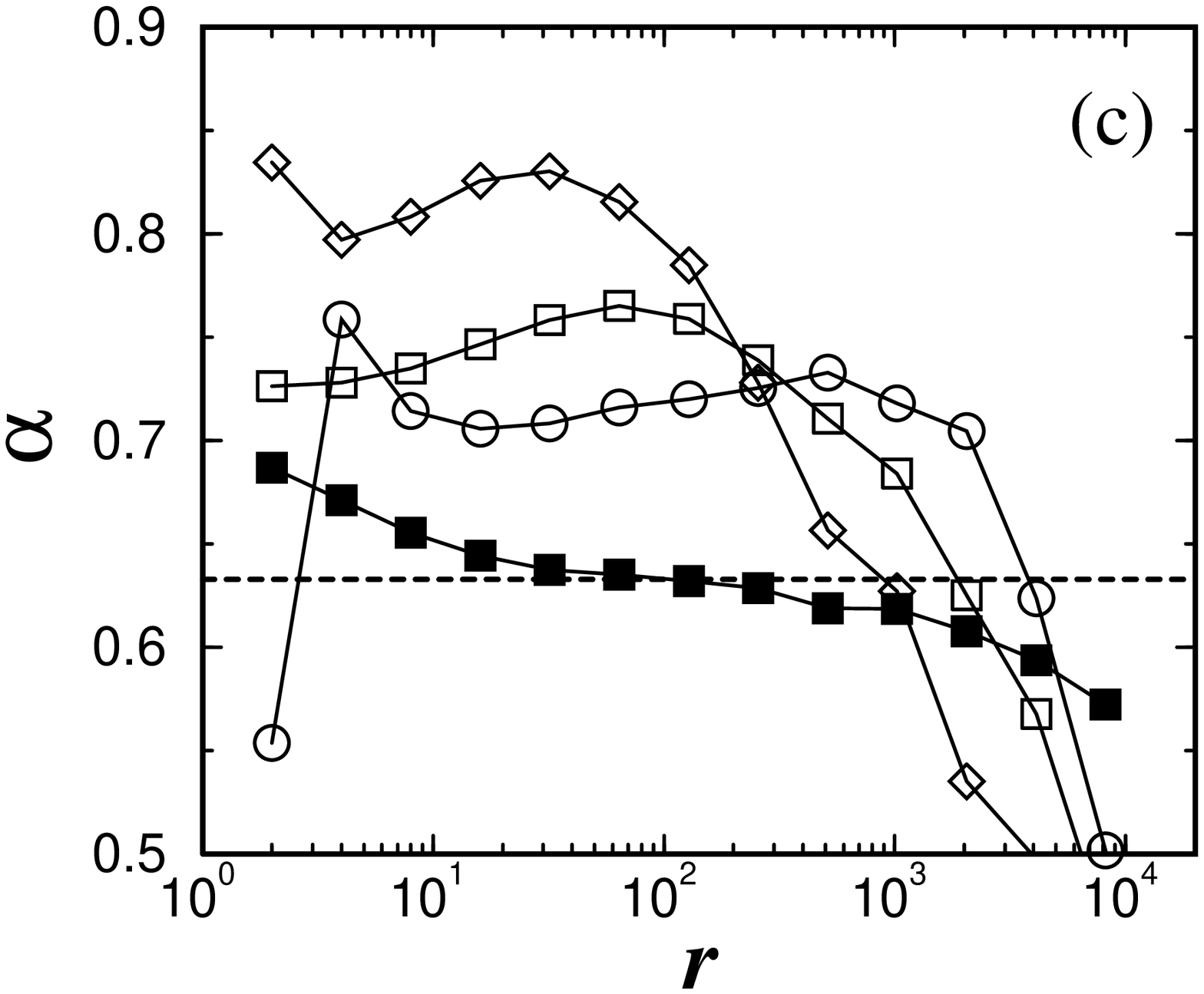}
\caption{
(a) 
Equal-time correlation function $C^2(r) \sim r^{2\alpha} $ for pinned
interfaces slightly below threshold: $F = 0.149$ ($L=65536$)
for Eq. (10) and $p=0.6623$ ($L=131072$) for the automaton. 
We averaged the data over 12 independent runs.
Both lines give $\alpha = 0.63 \pm 0.01$.
(b) Critical effective exponents 
$\alpha (r) = \log [C^2(r)/C^2(r/2)]/\log 4$
corresponding to (a) for Eq. (10) (full symbols).
The dashed line indicates the exponent $\alpha \simeq 0.633$
of the DPD model.
In addition, the effective exponents $\alpha _q (r)$ 
for different moments $q$   
of the 
equal-time correlation function 
$C^q(r)$
for moving interfaces above threshold 
($F=0.157$, $L=32768$) are shown: $q=1$ (open circles), $q=2$
(open squares), and $q=4$ (open diamonds).
The data were averaged over 30 disorder distributions so that 
the statistical uncertainties are of the size of the symbols 
or smaller.
(c) The same as (b) but for the automaton. 
The open symbols are the effective exponents for moving 
interfaces at $p=0.665$ (30 disorder distributions with $L=32768$).
}
\label{Cr}
\end{figure}

We measure $C^q(r,t)$ for the continuum 
equation (\ref{hdt}) and for the automaton and find results 
very similar to the DPD model \cite{pdp} (see Fig. \ref{Cr}b and 
\ref{Cr}c).
The effective exponents $\alpha _q (r)$ increase with $q$. The reason 
is that the moving regions (large slopes) have an increasing weight
with increasing $q$ \cite{pdp}.

Another possibility to investigate the scaling of moving interfaces
is a measurement of the height-height correlation function,
$ c^2(\tau)=\langle \overline {[h_{i}(t+\tau) -  h_i (t)]^2}   \rangle$,
for $t \gg \xi^z$ and $\xi < L$.
For $\tau \ll \xi ^z$, the height-height correlation
function scales as $c^2(\tau) \sim \tau ^{2 \beta}$, provided the interface 
is self-affine. As an illustration we consider the case $\lambda = 0$.

\subsubsection*{Height-height correlation function for the case $\lambda = 0$}

The growth exponent $\beta$ has been determined in Refs. \cite{gimo,MakAma}
by simulations of the automaton model  
Eqs. (\ref{f_i}), (\ref{h_i}) with $\lambda = 0$.
The scaling of the interface width at threshold, Eq. (\ref{w2t}),
yields 
$\beta \simeq 0.88$ \cite{gimo} and
$\beta \simeq 0.85$
\cite{MakAma}. Here, we measure the
effective exponents in the steady-state regime, 
$\beta(\tau)  =  \log [c^2(\tau) / c^2(\tau/2)] / \log 4$.

\begin{figure}
\epsfxsize=\linewidth
\epsfysize=0.27 \textheight
\epsfbox{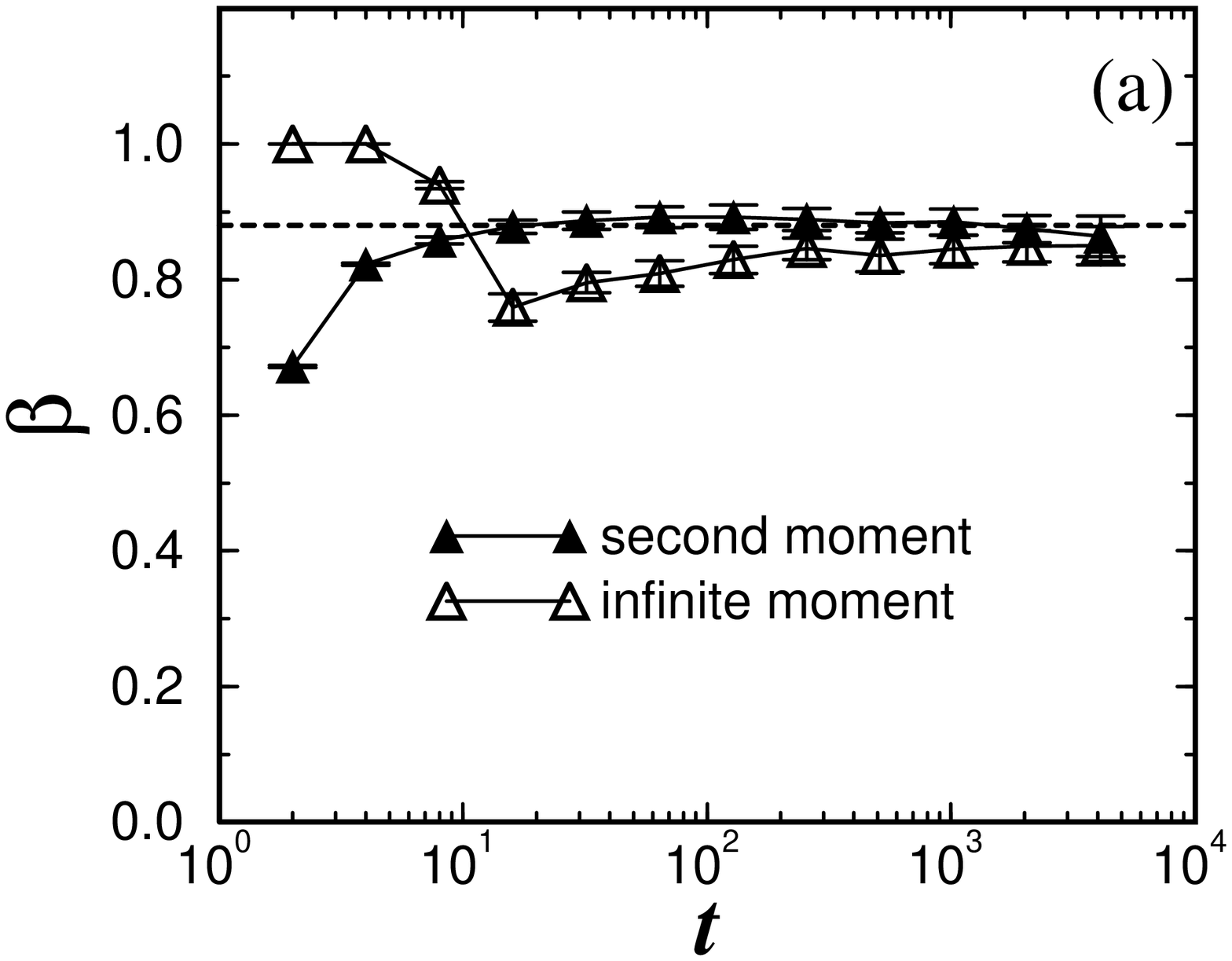}
\epsfxsize=\linewidth
\epsfysize=0.27 \textheight
\epsfbox{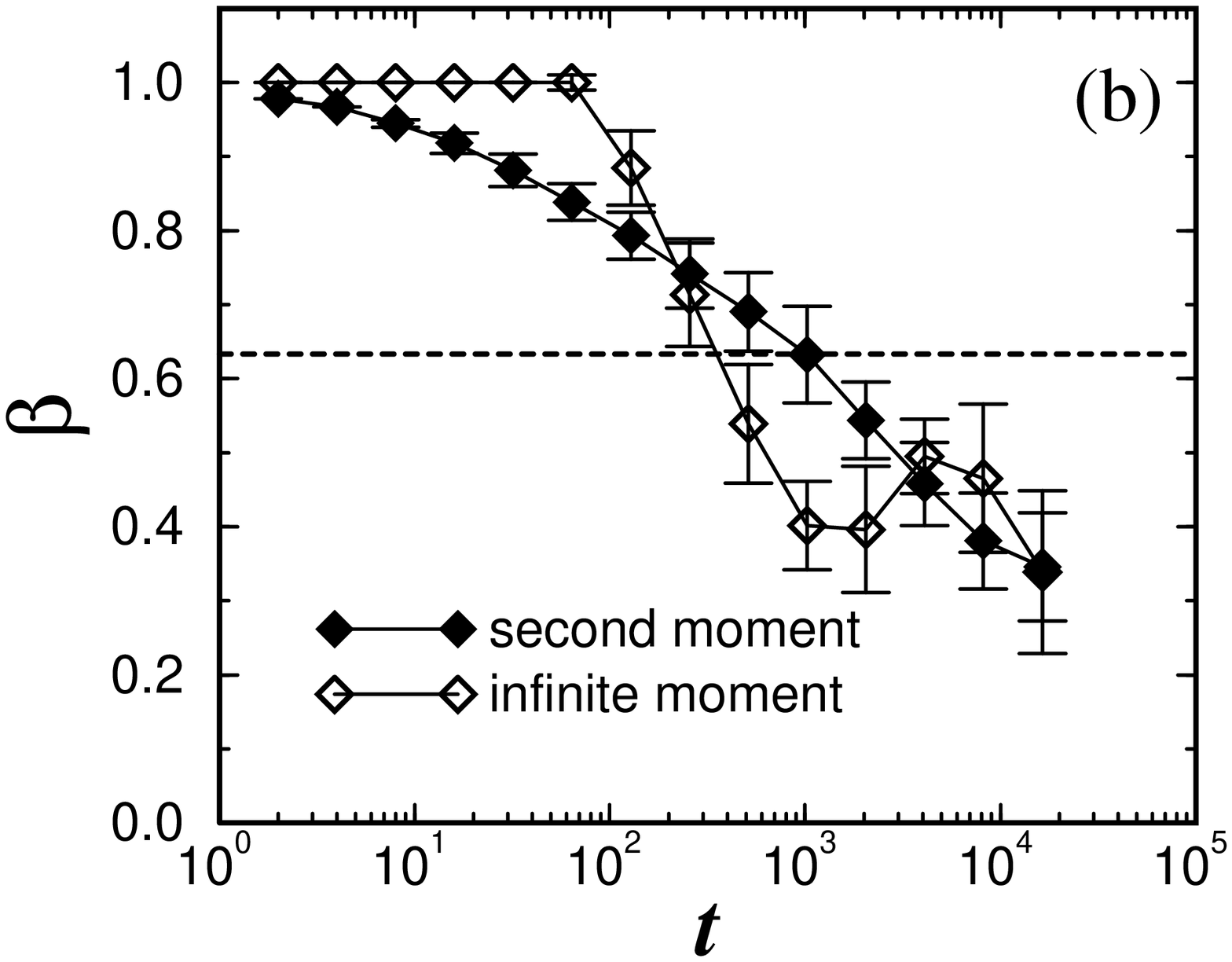}
\caption{
(a) Effective exponents 
$\beta(\tau) $ (full symbols) 
for the automaton with
$\lambda =0$, $\gamma =1$, $g=1$, $p=0.801>p_c\simeq 0.8004$, and
$L=32768$. The data were averaged over 30 disorder distributions.
The dashed line indicates the critical exponent
$\beta \simeq 0.88$ [22] 
from a measurement of the global width at threshold.
The effective exponents $\beta _\infty (\tau)$ 
(open symbols) seem to
approach this value for large $\tau$.
(b)
The same as (a) but for the DPD model of Ref. [17] with 
$p=0.467 > p_c \simeq 0.461$ (30 runs with $L=32768$). 
}
\label{betamfd}
\end{figure}

{}From Fig. \ref{betamfd}a we see that $\beta (\tau)$ is 
approximately constant over three orders of magnitude of $\tau$,
with $\beta = 0.88 \pm 0.01$. 
The effective exponent $\beta _\infty (\tau)$ of the infinite moment of 
the height-height correlation function, 
$c_\infty (\tau) = \langle \max _i \{h_i(t+\tau) - h_i(t) \} \rangle
\sim \tau ^{\beta _\infty}$, is also shown in Fig. \ref{betamfd}a. 
The effective exponent $\beta _\infty (\tau) $
increases slightly for $\tau \ge 16$
and approaches a value $\beta _\infty$ which is consistent with 
the values of Refs. \cite{gimo} and \cite{MakAma} obtained 
by the width at threshold. We conclude 
that the interface in  
Eq. (\ref{dht}) with $\lambda = 0$ is self-affine with 
the same growth exponent $\beta$ for both $F=F_c$ and $F>F_c$. 

\subsubsection*{Height-height correlation function for the anisotropic case}

In contrast, the same measurement for the DPD model 
(see Fig. \ref{betamfd}b) gives decreasing effective exponents
$\beta (\tau)$ and $\beta _\infty (\tau)$, which 
shows that there is no scaling with the critical growth exponent 
$\beta \simeq 0.63$. For small times $\tau$, the effective exponents 
$\beta (\tau)$ and $\beta _\infty (\tau)$ are 
roughly one, which can be understood from the lateral motion 
of the parts of the interface with linear slope.

\begin{figure}
\epsfxsize=\linewidth
\epsfysize=0.27 \textheight
\epsfbox{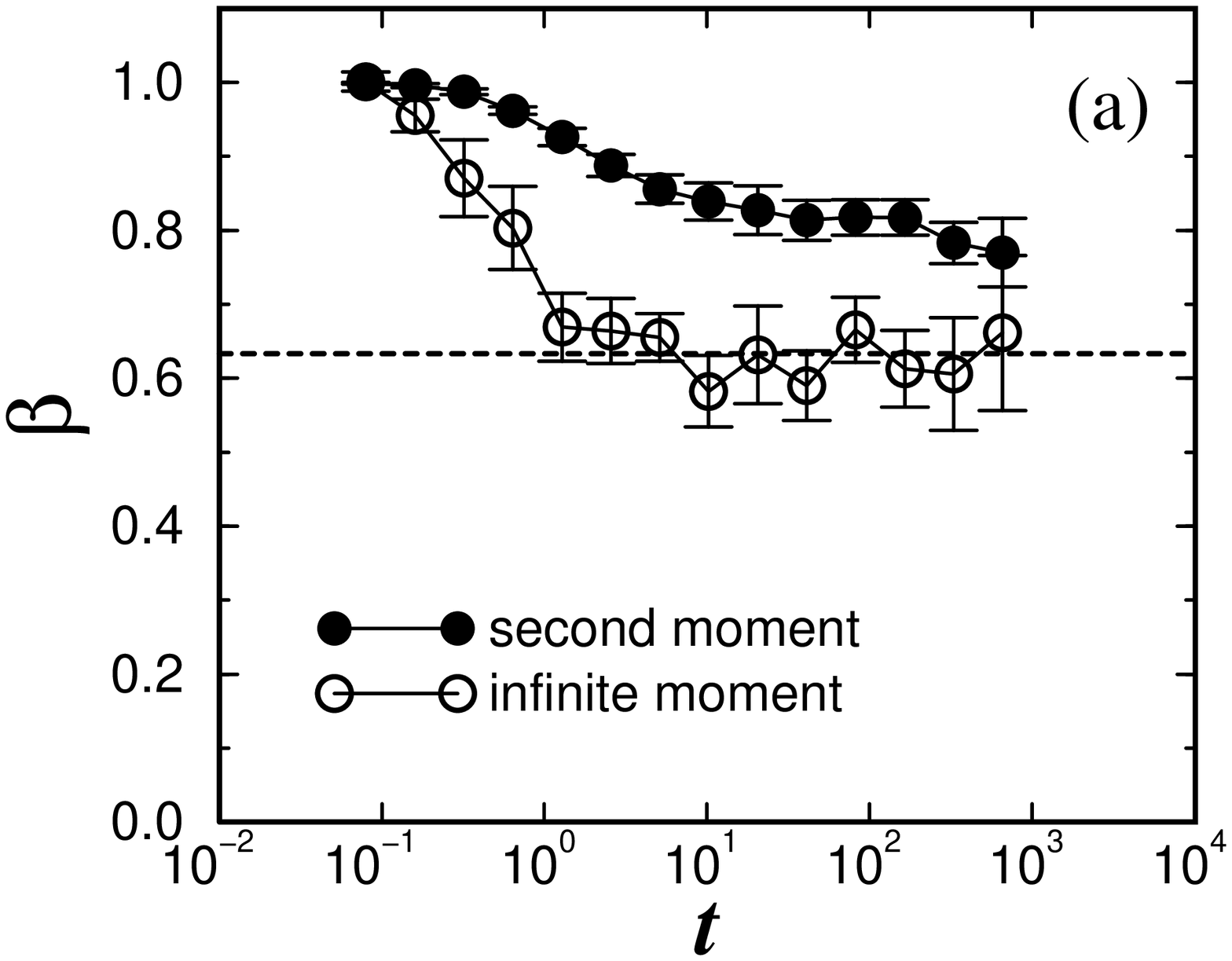}
\epsfxsize=\linewidth
\epsfysize=0.27 \textheight
\epsfbox{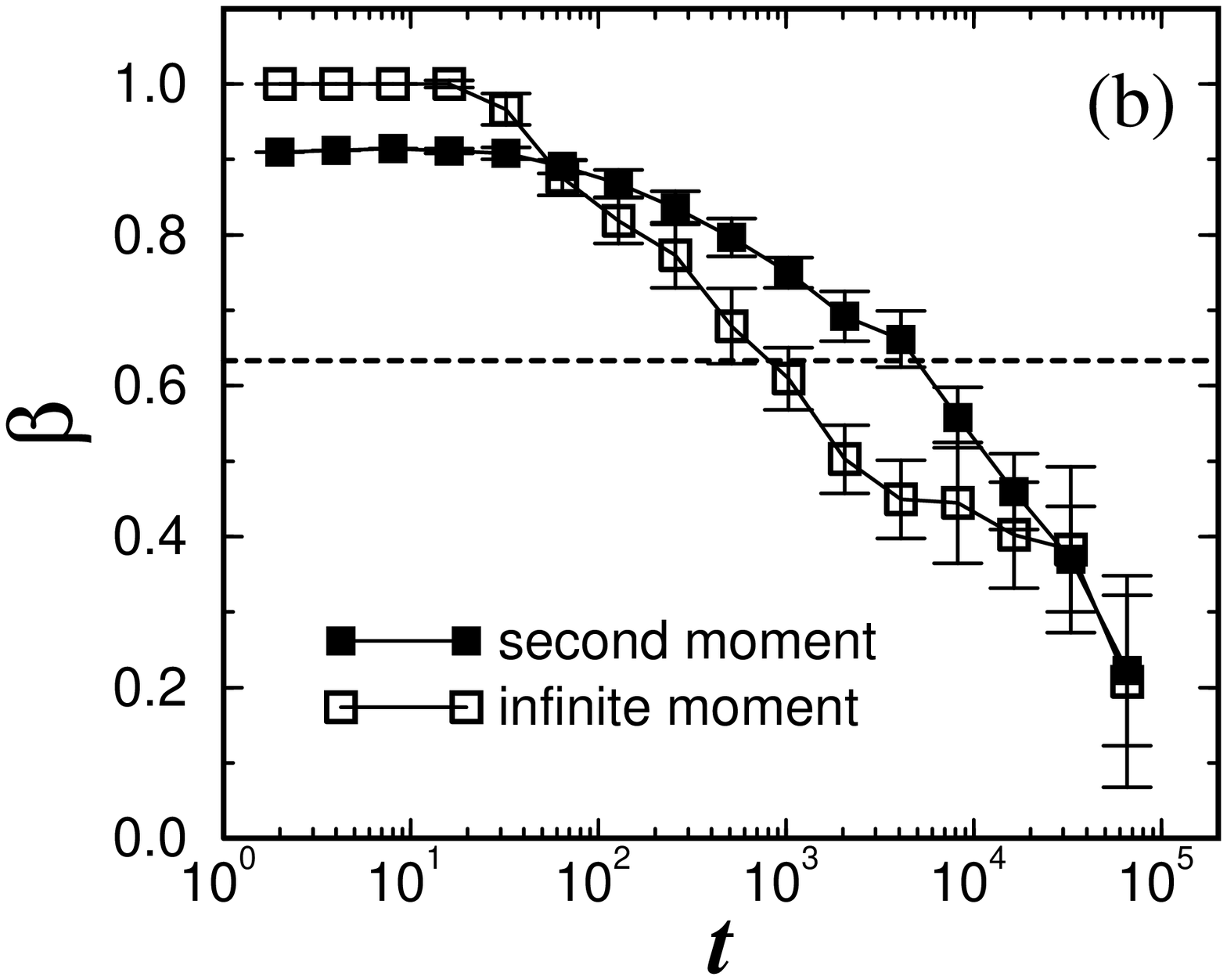}
\caption{
(a) 
Effective exponents $\beta (\tau)$ and $\beta _\infty (\tau)$
for the continuum equation 
($F=0.157 > F_c \simeq 0.1511$, 30 runs with $L=32768$).
(b)
The same as (a) but for the automaton 
($p=0.665 > p_c \simeq 0.6631$, 30 runs with $L=32768$).
}
\label{betamca}
\end{figure}

The behavior of the automaton with $\lambda >0$ 
is very similar (see Fig. \ref{betamca}b).
The effective exponent $\beta (\tau)$ for the continuum equation 
(Fig. \ref{betamca}a) 
also decreases with $\tau$ 
but shows a plateau at a value $\beta \simeq 0.82$.
The infinite moment $c_\infty (\tau)$ seems to show a scaling for 
large $\tau$ with $\beta _\infty \simeq 0.63$. It is not clear 
whether or not these plateaus in Fig. \ref{betamca}a can be
considered as some (multi-) scaling regime.
We can only conclude that the behavior is different from that 
of a self-affine interface which is characterized 
by the same critical growth exponent $\beta$ 
for $F=F_c$ and $F>F_c$ as long as 
time scales $\tau \ll \xi ^z$ are considered.

\subsection{Scaling of the correlation length and the velocity} 

We now proceed to determine the correlation length exponent $\nu$ 
and the velocity exponent $\theta$, 
which characterize 
the behavior of the interface close to the depinning transition.

\begin{figure}
\epsfxsize=\linewidth
\epsfysize=0.27 \textheight
\epsfbox{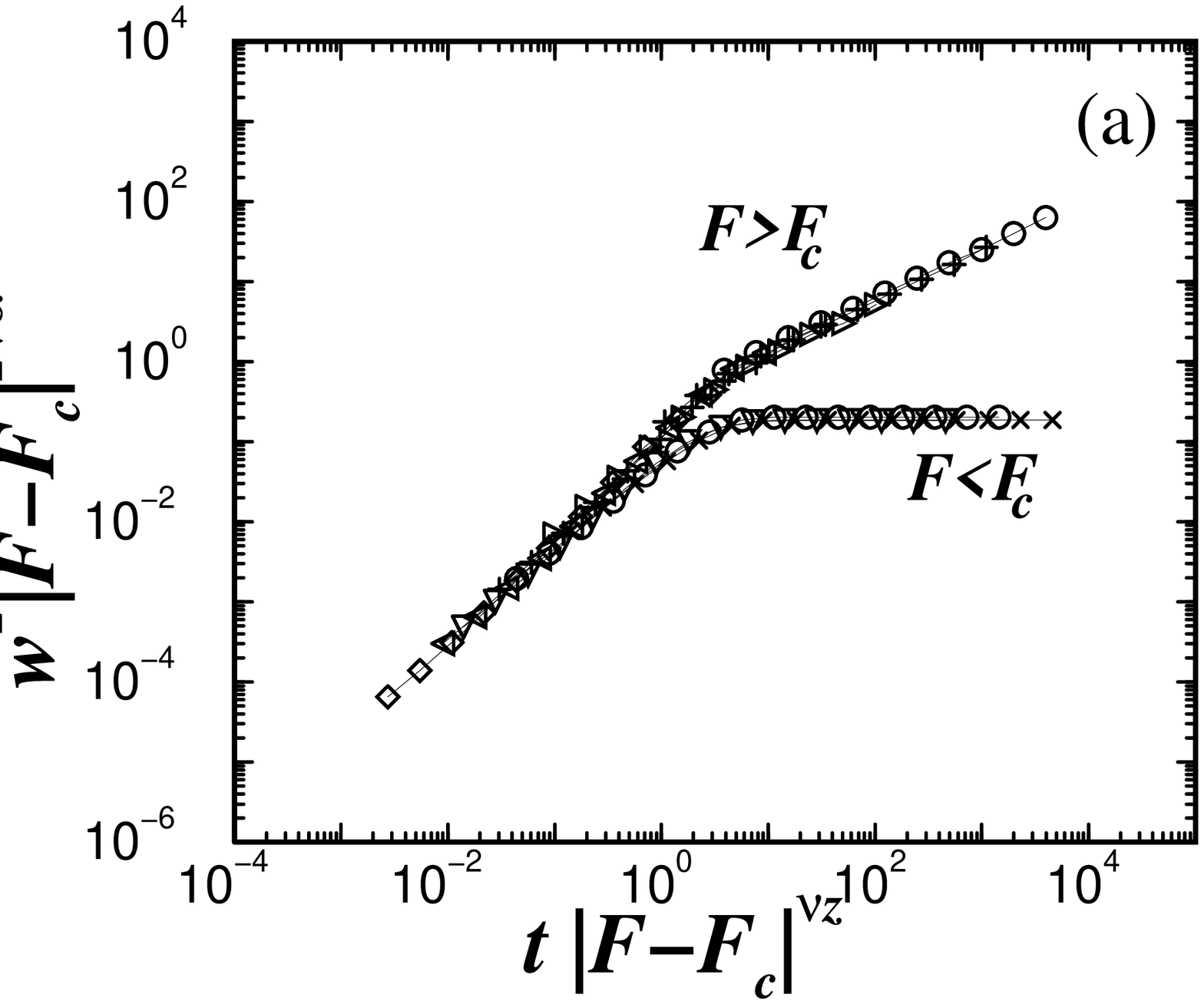}
\epsfxsize=\linewidth
\epsfysize=0.27 \textheight
\epsfbox{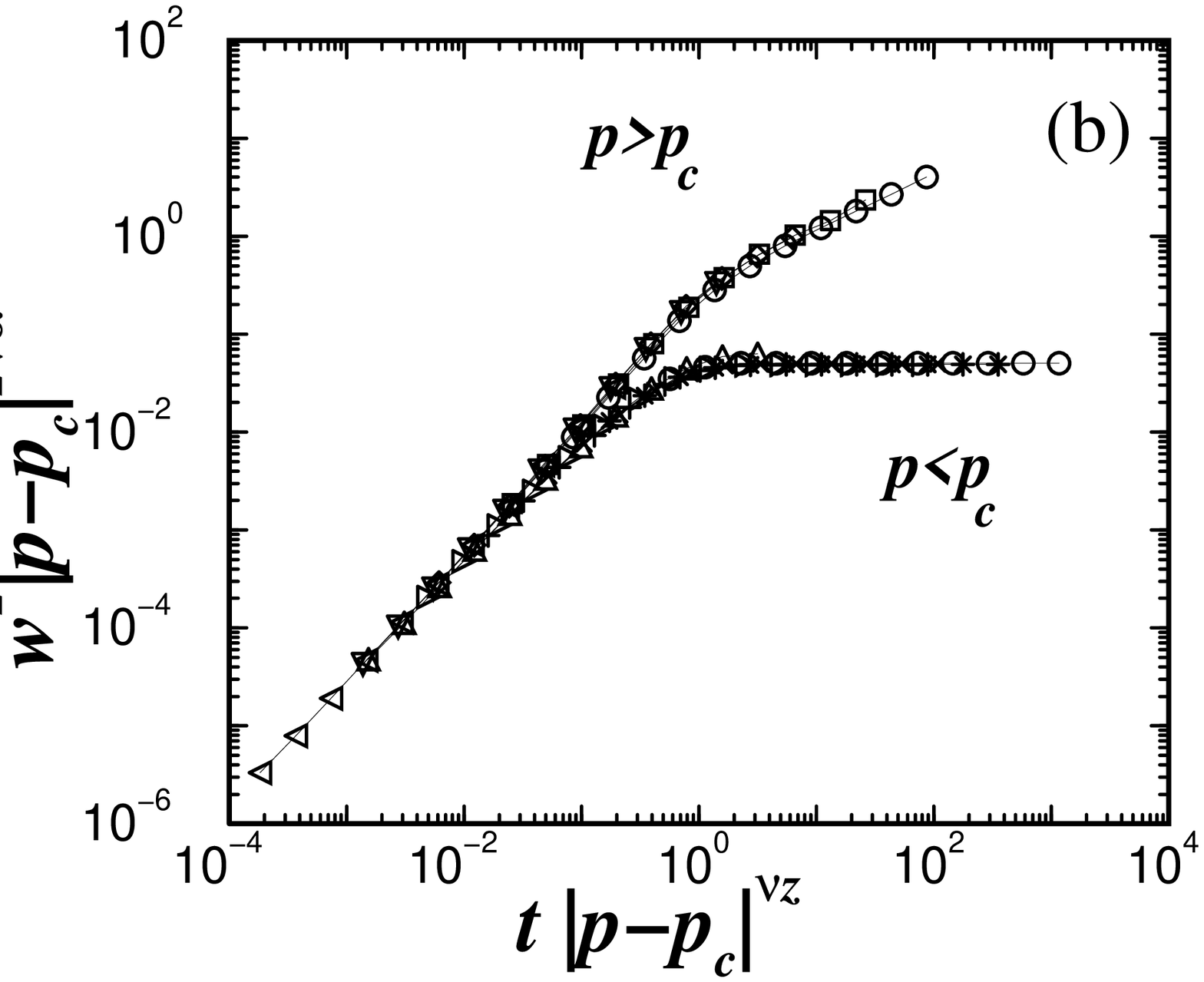}
\caption{
(a) Scaling plot according to Eq. (5) for the continuum equation.
The same plotting symbol is used for data at a given driving force $F$.
Above threshold,
$F$ is in the range $0.163 \le F \le 1$ and below threshold 
$0.03 \le F \le 0.12 $.
For each value of $F$ we simulated 50 independent runs with 
$L=16384$.
The best data collapse is achieved with 
$\alpha \simeq 0.63$, $z \simeq 1$, and $\nu \simeq 1.7$.
(b)
Scaling plot according to Eq. (5) for the automaton.
Above threshold,
$p$ is in the range $0.664 \le p \le 0.695$ (20 independent runs)
and below threshold 
$0.567 \le p \le  0.66 $ (100 independent runs).
The exponents for the best data collapse are 
$\alpha \simeq 0.63$, $z \simeq 1$, and $\nu \simeq 1.72$.
}
\label{nu}
\end{figure}

In Fig. \ref{nu},  scaling plots according to Eqs. (\ref{xi})
and (\ref{w^2sca})
are shown. Since we already determined the threshold and 
the critical exponents $\alpha$ and $z=\alpha / \beta$, 
we can tune the correlation length exponent $\nu$ to achieve the 
best data collapse. The result for Eq. (\ref{hdt}) is 
$\nu = 1.7 \pm 0.1$ and for the automaton $\nu = 1.72 \pm 0.03$.
The corrections to the scaling $w^2(\xi) \sim \xi ^{2 \alpha}$
due to finite size effects are much larger 
for the continuum equation. Therefore simulations
very close to the threshold are not shown in Fig. \ref{nu}a and the 
error bar on the result for $\nu$ is rather large. 
The results for the correlation length exponent 
are consistent with 
the prediction of directed percolation, $\nu \simeq 1.733$.

\begin{figure}
\epsfxsize=\linewidth
\epsfysize=0.27 \textheight
\epsfbox{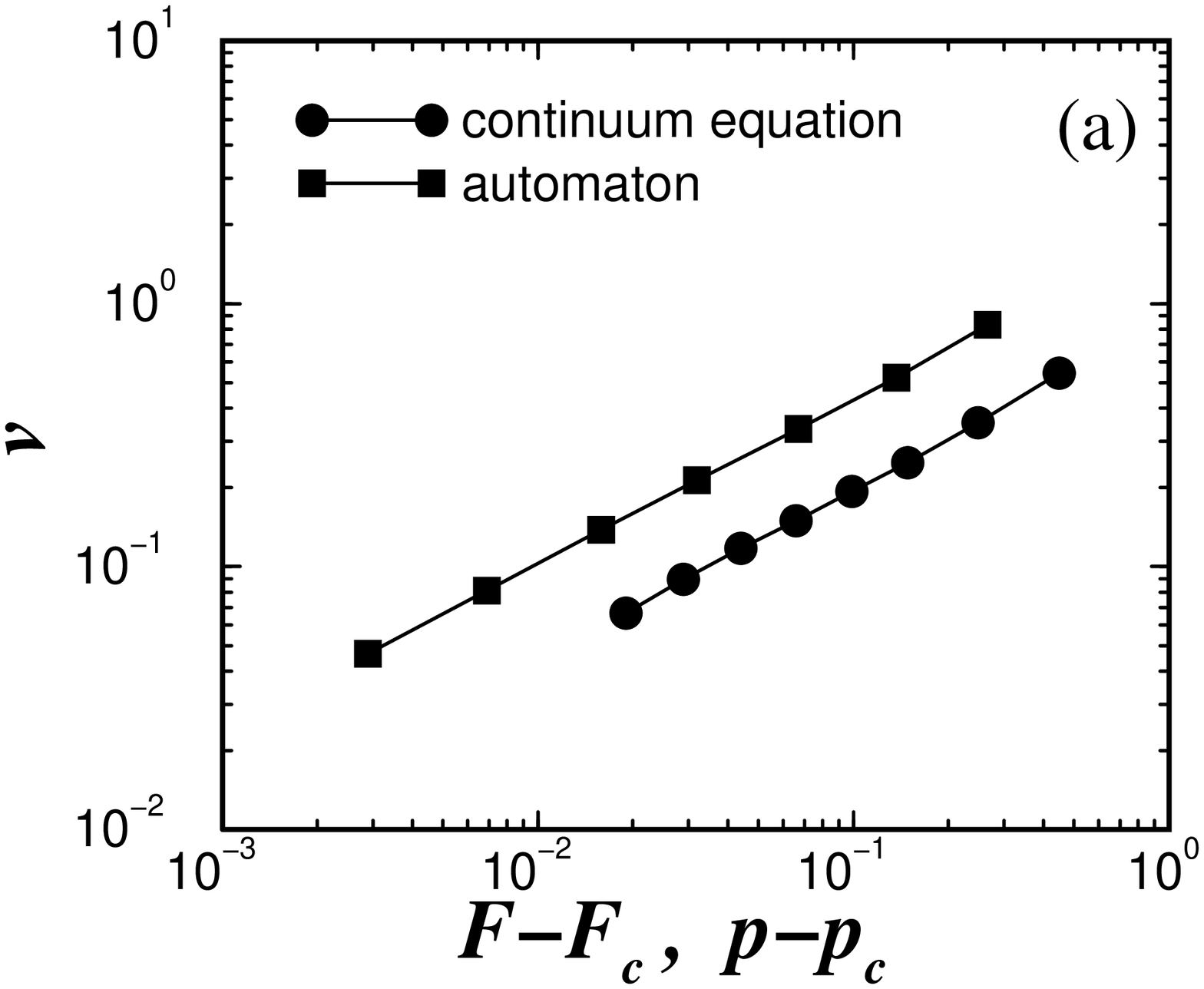}
\epsfxsize=\linewidth
\epsfysize=0.27 \textheight
\epsfbox{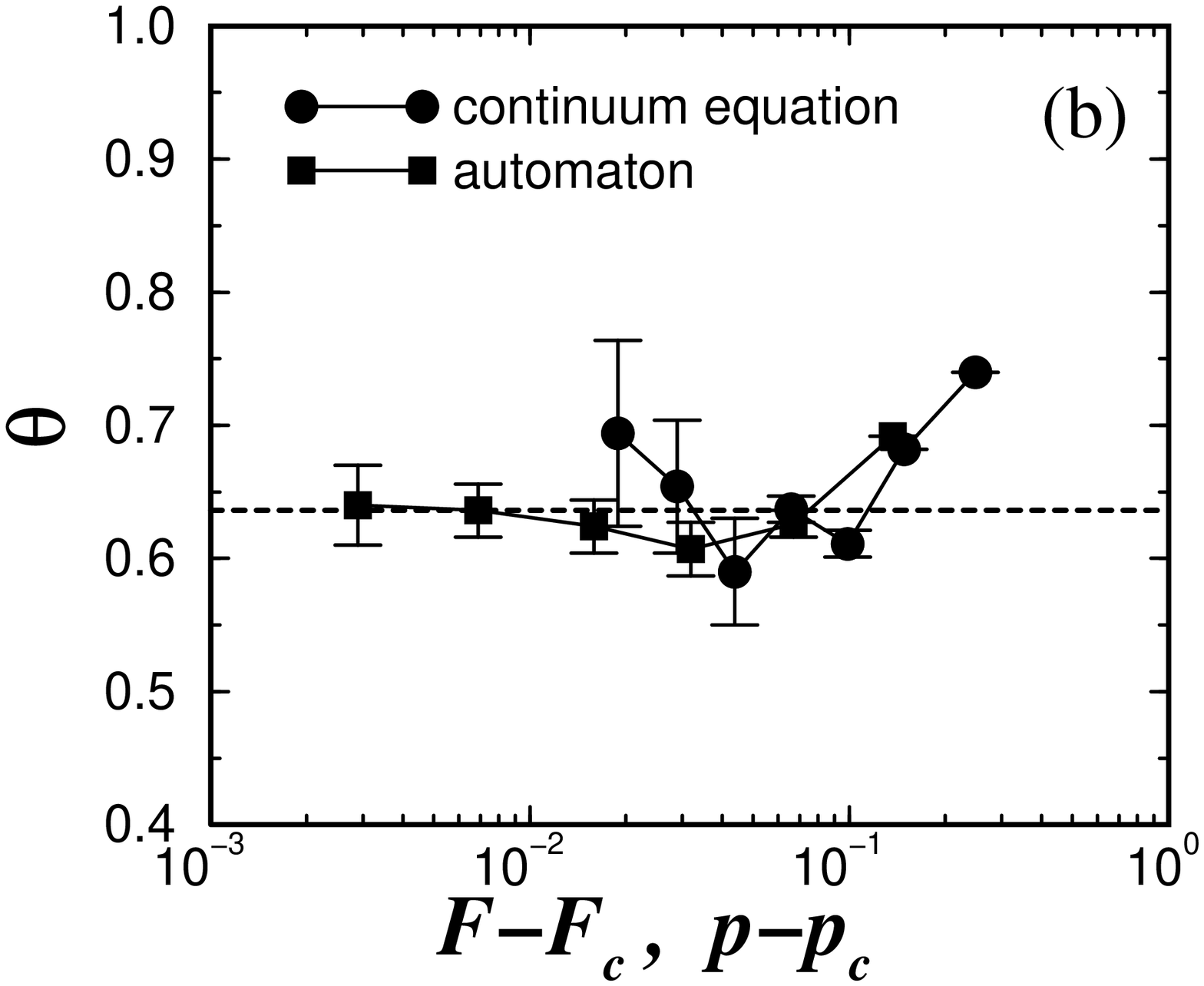}
\caption{
(a) The interface velocity $v$ as a function of the driving force.
The velocity is averaged over sufficiently long time intervals,
so that the statistical uncertainties are smaller than the size 
of the symbols.
For the continuum equation we obtain $\theta = 0.64 \pm 0.05$ 
($L=65536$) and 
for the automaton $\theta = 0.63 \pm 0.02$ ($L=131072$). 
(b) The effective exponents
$\theta (F-F_c) = \log [v(F)/v(F')] / \log [(F-F_c)/(F'-F_c)]$. 
The dashed line indicates
the critical value $\theta \simeq 0.636$ of the DPD model. 
}
\label{theta}
\end{figure}

The growth exponent $\beta _m $ defined in Eq. (\ref{w^2cas}) 
for driving forces $F>F_c$ and time scales $t \gg \xi ^z$ 
is found to be
$\beta _m = 0.32 \pm 0.03$ for the continuum equation 
and $\beta _m = 0.32\pm 0.02$ for the automaton. These values for 
$\beta _m$ are consistent with the exponent $\beta _m = 1/3$ 
of the KPZ equation \cite{KPZ}. This supports the
picture that the quenched disorder $\eta (x,h)$ crosses over to 
a thermal noise $\eta (x,vt)$ on time scales 
$t \gg \xi ^z$ (see Fig. \ref{nu}).

The scaling Eq. (\ref{v}) of the steady-state 
velocity $v$ and the corresponding effective
exponents $\theta$ are shown in Fig. \ref{theta}. The result
for the continuum equation is $\theta = 0.64 \pm 0.05$,
and $\theta = 0.63 \pm 0.02$ for the automaton, again
in agreement with the value of the DPD model, 
$\theta \simeq 0.636$ \cite{pdp}.

\section{Summary and Conclusions}

The results for the roughness of Eq. (\ref{dht}) with and without the 
term $\lambda (\nabla h)^2$ are summarized in Table I.
\begin{center}
\begin{tabular}{|c||c|c|}\hline
&Eq. (9), $\lambda =0$& Eq. (9), $\lambda > 0$  \\ \hline
$F=F_c$, $v \to 0$ & $\beta \simeq 0.88$  \cite{gimo} 
& $\alpha \simeq \beta \simeq 0.63$  \\
$F>F_c$, $v={\rm const.}$, $\tau \ll \xi ^z$& 
$\beta \simeq 0.88$ & not self-affine \\
$F>F_c$, $t \gg \xi ^z$
& $\beta _m= 1/4$ \cite{gimo}& $\beta _m= 1/3$\\ \hline
\end{tabular}
\end{center}
\vskip 0.2truecm
Table I. {Behavior of the roughness for Eq. (\ref{dht})
in $d=1$.} 
\vskip 0.4truecm

For $\lambda > 0$, pinned interfaces at $F=F_c$ are self-affine with 
a roughness exponent $\alpha \simeq 0.63$. Moving interfaces are not
self-affine
which we demonstrated by measuring the correlation functions
$C^q(r,t)$ and $c(\tau)$ (see Figs. 2-4).  
This is in comparison  
to the simpler behavior of 
Eq. (\ref{dht}) with $\lambda =0$, where not only pinned interfaces 
are self-affine (on length scales $l \ll \xi$) but also moving
interfaces on time scales $\tau \ll \xi ^z$.
The latter can be seen from 
the fact that 
the height-height correlation function $c(\tau)$
for an interface moving with constant velocity 
scales with the same 
growth exponent $\beta \simeq 0.88$ 
as the global width at $F=F_c$ 
(see Fig. 3a and Table I).
The values for the exponent $\beta _m$  
(Eq. (\ref{w^2cas}))
are obtained by the scaling of the global width $w(t)$ on time 
scales $t \gg \xi ^z$ and 
correspond to those of the Edwards-Wilkinson equation ($\lambda =0$)
\cite {EW} and the KPZ equation ($\lambda > 0$) \cite{KPZ},
respectively.

The results for the critical exponents characterizing the
depinning transition of Eq. (\ref{dht}) with $\lambda > 0$ are summarized in  
Table II.
\begin{center}
\begin{tabular}{|c||c|c|c|}\hline
\bf Exponent &\bf DPD &\bf continuum eq. &\bf automaton \\ \hline
roughness, $\alpha$ & 0.633  & 0.63 $\pm$ 0.01 & 0.63 $\pm$ 0.01  \\
growth, $\beta$ & 0.633 & 0.64 $\pm$ 0.02 & 0.64 $\pm$ 0.02  \\
correlation length, $\nu$ & 1.733 & 1.7 $\pm$ 0.1 & 1.72 $\pm$ 0.03 \\
velocity, $\theta$ & 0.636 & 0.64 $\pm$ 0.05 & 0.63 $\pm$ 0.02 \\ \hline
\end{tabular}
\end{center}
\vskip 0.2truecm
Table II. {Comparison of the critical exponents of the DPD model
with our numerical results.}
\vskip 0.4truecm

The numerical results for both, the continuum equation (\ref{hdt}) and 
the automaton are in excellent agreement with the 
DPD model. We therefore expect that the critical exponents
of the anisotropic depinning model Eq. (\ref{dht}) are exactly
given by the exponents of directed percolation. 

In the DPD model, the dynamic and the static behavior is determined
by directed percolation paths of pinning sites. Due to the 
restricted solid-on-solid condition of the directed percolation
paths, the pinned regions of the interface have small gradients
(bounded by a slope one) \cite{pdp}. In contrast, the laterally moving 
regions have a linear slope of about two \cite{pdp}. This behavior is
analogous to that of Eq. (\ref{dht});
regions of the interface with large slopes are likely to move due to
the positive contribution of the term 
$\lambda (\nabla h)^2$ to the driving force. 
Regions of the interface with small gradients, the other hand, 
are easier to pin, due to the smaller contribution of 
$\lambda (\nabla h)^2$. 

The observation that the interface motion is mainly
due to the gradient term 
$\lambda (\nabla h)^2$, causes a clustering of growth sites, which
can be understood as follows.
A motion of an interface element $h(x) \to h(x) + dh$ 
increases the contribution of $\lambda (\nabla h)^2$
to the local force at $h(x + dx)$ or $h(x - dx)$.
Thus, this neighboring interface element is likely to be the next 
new growth site, resulting in a cluster 
of growth sites to be formed. The moving regions 
have larger slopes than the pinned parts.
As a consequence, a moving interface is not self-affine.
The clustering of growth sites in Eq. (\ref{dht}) does not destroy 
the dynamical scaling of global quantities of the interface,
such as the width and the velocity (see Eqs. (\ref{v}--\ref{w2t})).
The scaling relation (\ref{exasca}) is also fulfilled by the 
exponents of the DPD model and by our numerical results.

The behavior of moving interfaces  
deserves further investigation to understand better the 
concepts of scaling and self-affinity (see Sec. IV B).
Further, it would be very interesting to construct a
rigorous proof that Eq. (\ref{dht}) and the DPD model are in the
same universality class.

\section*{Acknowledgments}

I would like to thank 
L. A. N. Amaral, 
P. Bak,
S. V. Buldyrev, 
R. Cuerno,
S. Galluccio, 
S. T. Harrington, 
S. Havlin,
K. B. Lauritsen, 
H. A. Makse,
M. Paczuski,
H. E. Stanley,
L.-H. Tang, and 
S. Zapperi
for valuable discussions.
I acknowledge support from the Deutsche Forschungsgemeinschaft. 
The Center for Polymer Studies is supported by the 
National Science Foundation.

\end{multicols}

\end{document}